\documentclass[journal]{IEEEtran}
\ifCLASSINFOpdf
   \usepackage[pdftex]{graphicx}
\else
   \usepackage[dvips]{graphicx}
\fi

\usepackage{ulem} 
\usepackage[colorlinks=true, allcolors=blue]{hyperref}

\begin{document}

\title{120 km single-photon Brillouin optical time domain reflectometry}

\author{\IEEEauthorblockN{Maxime~Romanet\IEEEauthorrefmark{1},
        Luis~Miguel~Giraldo\IEEEauthorrefmark{1},
        Maxime~Zerbib\IEEEauthorrefmark{1},
        Etienne~Rochat\IEEEauthorrefmark{2},
        Kien~Phan~Huy\IEEEauthorrefmark{1} \IEEEauthorrefmark{3},
        and~Jean-Charles~Beugnot\IEEEauthorrefmark{1}}\\
\IEEEauthorblockA{\IEEEauthorrefmark{1}FEMTO-ST Institute, CNRS UMR 6174, Université de Franche-Comté, Besançon, France.}\\
\IEEEauthorblockA{\IEEEauthorrefmark{2}Omnisens S.A., Morges, Switzerland} \\
\IEEEauthorblockA{\IEEEauthorrefmark{3}FEMTO-ST Institute, CNRS UMR 6174, ENSMM, Besançon, France}
\thanks{Maxime Romanet is with the Department
of Optics, Institut FEMTO-ST, Besançon, France e-mail: maxime.romanet@femto-st.fr).}

\thanks{Manuscript received \today .}}

\maketitle

\begin{abstract}
We present a novel distributed Brillouin optical time domain reflectometer (BOTDR) using standard telecommunication fibers based on single-photon avalanche diodes (SPADs) in gated mode, $\nu -$BOTDR, with a range of 120~km and 10~m spatial resolution. We experimentally demonstrate the ability to perform a distributed temperature measurement, by detecting a hot spot at 100~km. Instead of using a frequency scan like conventional BOTDR, we use a frequency discriminator based on the slope of a fiber Bragg grating (FBG) to convert the count rate of the SPAD into a frequency shift.
A procedure to take into account the FBG drift during the acquisition and perform sensitive and reliable distributed measurements is described. We also present the possibility to differentiate strain and temperature. 
\end{abstract}


\IEEEpeerreviewmaketitle

\section{Introduction}
\IEEEPARstart{B}{rillouin} scattering based distributed optical fiber sensors have become essential tools for civil engineering applications  such as monitoring of pipelines or submarine cables \cite{motil2016state,hartog_introduction_2017}. Their field of application has expanded opening up new horizons for geophysicists and natural scientists \cite{wang_test_2009,yin_real-time_2010,zeni_brillouin_2015}. These sensors, which take advantage of the frequency dependence of Brillouin gain on temperature and strain, fall into two categories \cite{hartog_introduction_2017,bai_recent_2019}. Brillouin optical time domain analysis sensors (BOTDA) use stimulated Brillouin scattering and require access to both ends of the fiber. Brillouin optical time domain reflectometers (BOTDR) that measure spontaneous Brillouin scattering require access to only one end of the fiber and provide a clear advantage in terms of simplicity. Increasing signal-to-noise ratio (SNR) represents a major challenge for these sensors. It determines the range of the sensor, typically 75 km on commercial devices. Numerous improvements have been proposed in terms of probe light \cite{li_snr_2012,zhang_performance_2014}, remote amplification with Raman \cite{cho_50-km_2003,cho_enhanced_2004,song_100_2015} or EDFA amplifiers \cite{gyger2014extending,clement_enhancement_2021} or coding strategies \cite{soto_analysis_2008,soto_brillouin-based_2008,wang_enhancing_2017}. However, the specifications of the photodetector always limit the ultimate performance. With the development of quantum technologies, new detectors such as single-photon avalanche diodes (SPAD) and superconducting nanowire single-photon detector (SNSPD) become available \cite{eraerds_photon_2010}.  They have shown their efficiency in many OTDR applications, improving sensing range while maintaining a good resolution \cite{diamanti_15_2006,antony_single-photon_2012,hu_photon-counting_2012,zhao_long-haul_2015}.

However, few studies have been carried out on single-photon Brillouin optical time domain reflectometers ($\nu-$BOTDR). They use either the Landau-Placzek ratio (LPR) requiring Rayleigh and Brillouin intensity measurements, to determine the temperature \cite{xia_brillouin_2016} or a scanning Fabry-Perot interferometer to scan the Brillouin resonance \cite{xia_distributed_2016}. Measurements were reported up to 50~km with an estimated  range of 90~km \cite{xia_distributed_2016}.

In this paper, we propose a single-photon $\nu$-BOTDR sensor to perform distributed temperature measurements using a narrow fiber Bragg grating (FBG) without frequency scan. This idea has already been used in schemes using traditional detectors with Mach-Zehnder interferometers as discriminators \cite{masoudi_distributed_2013}. In particular, our configuration allows any instability issue to be addressed with a simple post-processing of the data. 

In the first section, we will present our experimental setup and demonstrate its detection range. Then, we will explain the principle of temperature measurement with a narrow FBG used as a frequency discriminator and show our experimental results. In the third section, we will show how instabilities can be corrected through real-time monitoring and post-processing of the data, thus eliminating the need for a feedback loop. Finally, we will give an overview of how our setup can deal with temperature-strain cross-sensitivity. 

\section{Measurement setup and sensing range}

\begin{figure}[!t]
\centering
\includegraphics[width=3.5 in]{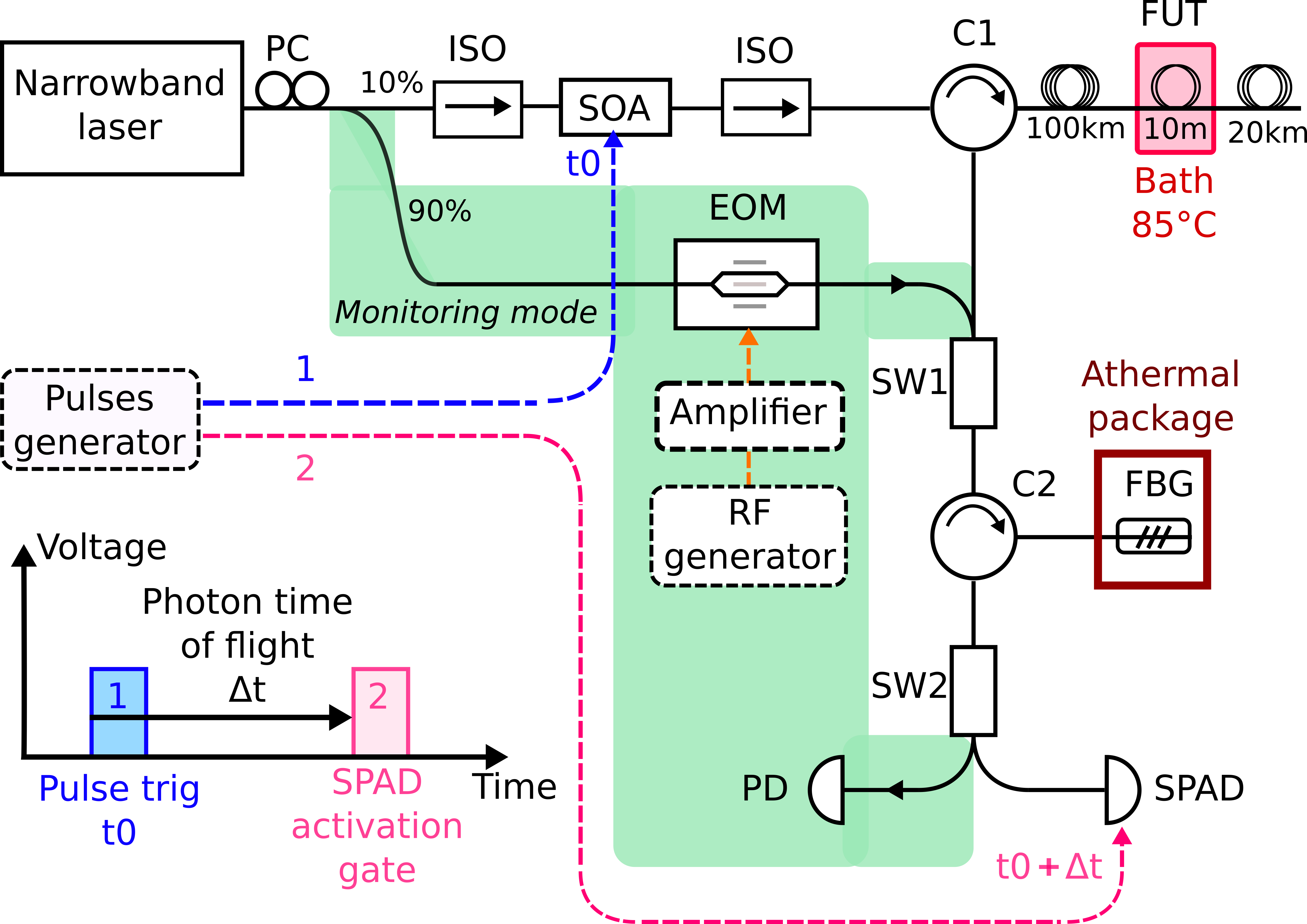}
\caption{$\nu-$BOTDR setup. Green area: FBG frequency response monitoring. PC: Polarization controller, ISO: Optical isolator, SOA: Semiconductor Optical Amplifier, EOM: Intensity Electro-Optic modulator, C1-C2: circulators, FUT: Fiber under test. Red area: Temperature controller bath (T = 85 °C). SW1-SW2: Optical switches, FBG: athermal packaged Fiber Bragg Grating, SPAD: Single-Photon Avalanche Diode, PD: Photodiode. Dash line: electrical cables. }
\label{fig:setup}
\end{figure}

Our experimental setup is shown in Fig.~\ref{fig:setup}. A tunable narrowband laser (Apex technologies)  sends light signal at $\lambda=1547$~nm into a polarization controller (PC). The light is split in two beams by a 90/10 coupler. The beam with 10\% of the light reaches an optical isolator (ISO) and a semiconductor optical amplifier (SOA) that shapes the pump pulses. Pulses go through a second optical isolator and an optical circulator (C1) before entering the fiber under test (FUT): a 120~km single-mode fiber (G625D). \\
Within the FUT, the light experiences spontaneous Brillouin scattering. It is scattered by thermally excited phonons that propagate along the fiber. The back-scattered light is shifted in frequency by the Doppler effect. The frequency shift is related to the acoustic velocity in the SMF, which depends on the temperature and the strain. The back-scattered light returns to the circulator C1 where it is routed towards an optical switch SW1. In \textit{measurement mode}, the backscattered light is sent through SW1 to a circulator C2 and a narrow Fiber-Bragg Grating (FBG) that serves as a frequency discriminating reflector. Then the filtered backscattered light reaches a second switch SW2. SW2 carries the backscattered photons to an infrared SPAD from Aur\'ea Technology operated in the so-called gated mode or Geiger mode. The applied bias is above the breakdown voltage, in order to have sufficient gain to be able to detect a single incident photon. The SPAD is based on InGaAs/InP, cooled around -50°C. 

In order to make a distributed measurement along the optical fiber, it is important to locate precisely the position at which the light was backscattered; this is achieved by a time-of-flight measurement. As shown in Fig.~\ref{fig:setup}, by controlling the time delay $\Delta t$ between the launch of the pulse from the SOA (blue 1) and the activation gate of the SPAD (pink 2), we can carry out measurements at a particular distance $d$: 

\begin{equation}
\centering 
\Delta_t = \frac{2 d}{c} {n_{SMF}} + t_{setup}
\label{eq:01}
\end{equation}

where c is the speed of light in vacuum, $n_{SMF}$ is the fiber refractive index, $d$ is the sensing distance, $t_{setup}$ is the delay due to propagation time in setup components (isolator, circulators, switch, patch cords, filter) and $\Delta t$ is the photon round trip time of flight in the fiber under test (FUT).

The main advantage of using SPADs over conventional detectors is to increase the sensing range of the distributed sensor, which is determined by the signal-to-noise ratio (SNR) of the whole setup. The signal strength is determined by several factors such as peak pulse power, propagation and component losses, Brillouin scattering efficiency and SPAD detection efficiency. The noise is determined by the extinction ratio of the pulses shaped by the SOA, the isolation/cross talk of the circulators C1 and C2, the filtering of the Rayleigh and Raman backscattering, the detection noise (darkcount, after pulsing...) and its management (dead-time). All these data are reported in table \ref{table:specifications}. Careful attention has been given to the setup to provide the best SNR with the available components. 

\begin{table}[ht!]
\renewcommand{\arraystretch}{1.3}
\caption{Component specifications. \\
$*$ Measured, $**$ Given by manufacturer test report.}
\label{table:specifications}
\centering
\begin{tabular}{|l|l|c|}
\hline
 & Emission wavelength for measurements & 1547.8~nm \\
Laser  & Tunability range & 1550 $\pm$ 20~nm\\
 & Tunability resolution & 1~pm\\
& Output power * & +11~dBm\\
\hline
& Input power * & 0~dBm\\
& Pulse peak power, $P_{peak}$ * & +15~dBm\\
SOA & Pulse extinction ratio * & 62~dB\\
& Pulse width * & 100~ns\\
& Pulse period, $T$ * & 1.25~ms\\
\hline
Circulator& C1 isolation/crosstalk * & 62~dB\\
& C2 isolation/crosstalk * & 51~dB\\
\hline
Fiber & G.652.D optical fiber, $\alpha_{SMF}$ * & 0.19~dB/km\\
\hline
 & Insertion losses, $\alpha_{FBG}$ * & 2.5~dB\\
FBG &  Line width (FWHM) * & 2.7~GHz \\
 & Attenuation after 9~GHz FBG center * & 42~dB \\

\hline
& Efficiency, $\eta_s$ * &  20~\% \\
& Dark count, $D_k$ (free Running) ** & $\le$ 1150~cps\\
SPAD & Dead-time ** &  20~$\mu$s \\
& Gate, $\tau$ * & 50~ns\\
& Dark count $d_k$ in our gated setup * &  0.0437~cps\\
\hline

\hline
\end{tabular}
\end{table}

Using the specifications in Table \ref{table:specifications}, we can predict the detection rate 

\begin{equation}
\centering
C_{,cps} =  \eta_{s} \frac{ \tau}{T} \frac{S_{B,W} }{\hbar \omega} \quad ,
\label{eq:02}
\end{equation}

where $\hbar$ is the reduced Planck constant, $\omega$ is the pulsation of light, $\eta_s$ the SPAD efficiency, $\tau$ the gate duration of the SPAD, $T$ the period and $S_{B,mW}$ the backscattered Brillouin power [mW] defined as:

\begin{equation}
\centering 
S_{B,mW} = 10^{\frac{S_{B,dBm}}{10}}, \quad [\mathrm{mW}]
\label{eq:03}
\end{equation}

The backscattered Brillouin power along the fiber is defined as:
\begin{equation}
\centering 
S_{B,dBm} = P_{\mathrm{peak}} -g_{B} -2 \,d\,\alpha_{SMF} -\alpha_{setup}, \quad [\mathrm{dBm}]
\label{eq:04}
\end{equation}

 where $P_{\mathrm{Peak}}$ is the pulse peak power injected in the fiber, $g_B$ is the Brillouin efficiency (78~dB in our configuration), $d$ is the probed distance [km], $\alpha_{SMF}$ are fiber losses and $\alpha_{{setup}}$ are components losses of the setup (circulators, filter, switch). \\

\begin{figure}[ht!]
\centering
\includegraphics[width=3.5 in]{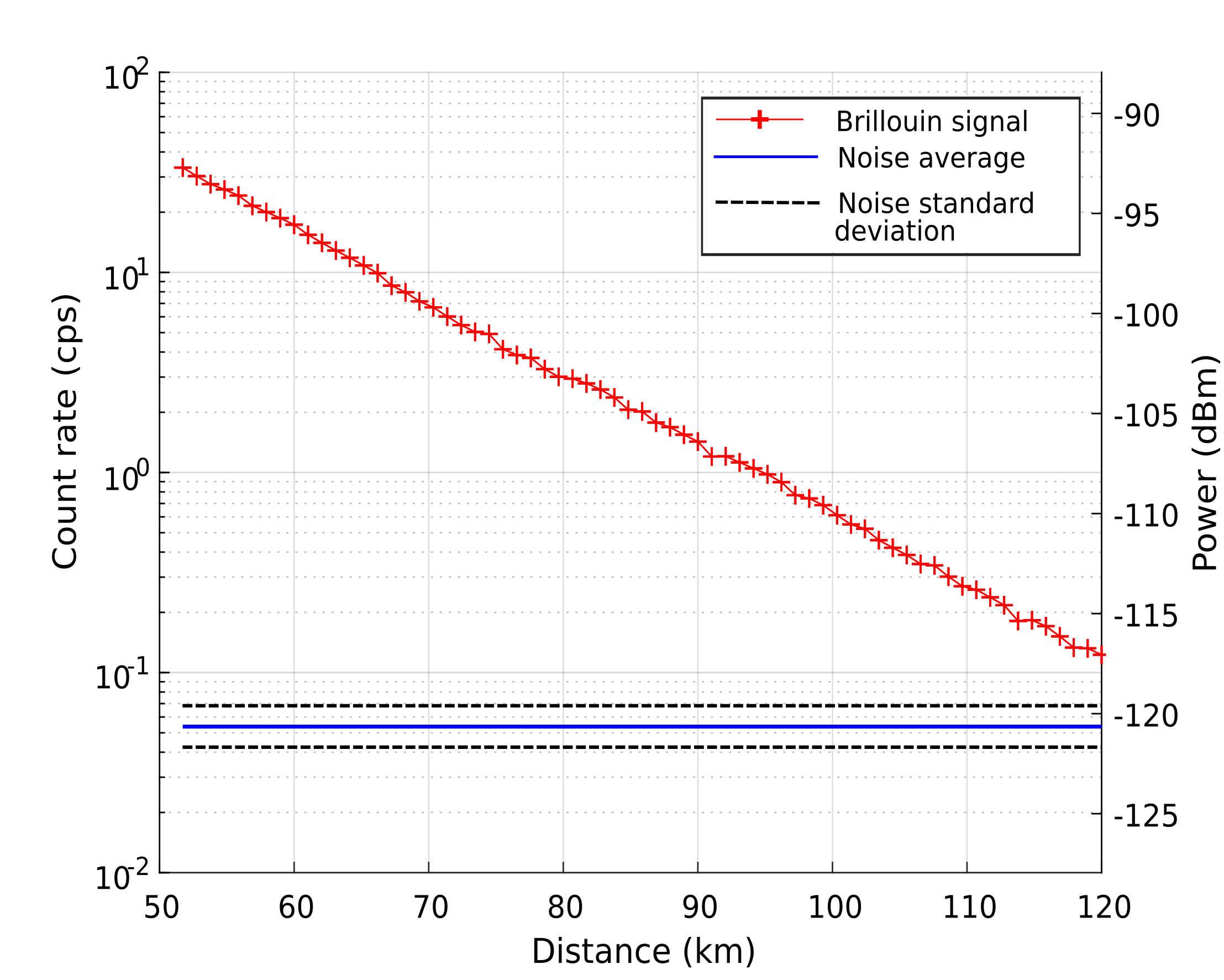}
\caption{(Red) $\nu$-BOTDR trace from 50~km to 120~km realized with operating point A (see Fig.~\ref{fig:FBG}), with a 10 m spatial resolution (SOA 100~ns pulse width), 50~ns SPAD gates. (Blue) Noise floor average. (Dark dotted) Noise floor standard deviation.}
\label{fig:Range}
\end{figure}

The experimental results are presented in Fig.~\ref{fig:Range}, 
the Brillouin anti-Stokes count rate is reported in count per second (cps) with a logarithmic scale as a function of the probed distance. 

 The low noise level of the sensor (0.05~cps), represented in blue on Fig.~\ref{fig:Range} allows us to reach a distance up to 120~km, with a 10~m spatial resolution, which is 30~km longer than the results reported in \cite{xia_distributed_2016}. On Fig.~\ref{fig:Range} we only measure from 50~km in order to avoid the nonlinear threshold of the SPAD (160~cps in our configuration). One solution to perform measurements before 50~km consists in adding an optical attenuator before the SPAD, or modify the setup parameters by reducing the optical pulse width or the gates width on the SPAD. This solution will increase the dynamic range of the $\nu-$BOTDR up to 45~dB. These results could be further improved by using a more efficient single-photon detector such as a SNSPD with the disadvantage of using a device cooled to a few Kelvins, which is out of the scope of this article. At 120~km, the sensor still has a signal-to-noise ratio (SNR) of 3.6~dB, which demonstrates the extended range of the $\nu-$BOTDR.

\section{Temperature measurements}
BOTDR sensors use the Brillouin frequency shift, to measure the temperature along an optical fiber. Usually, a heterodyne setup with a local oscillator is used. Such a configuration is not allowed in our case because the local oscillator would blind our single-photon detector. 

\begin{figure}[!t]
\centering 
\includegraphics[width=3.5 in]{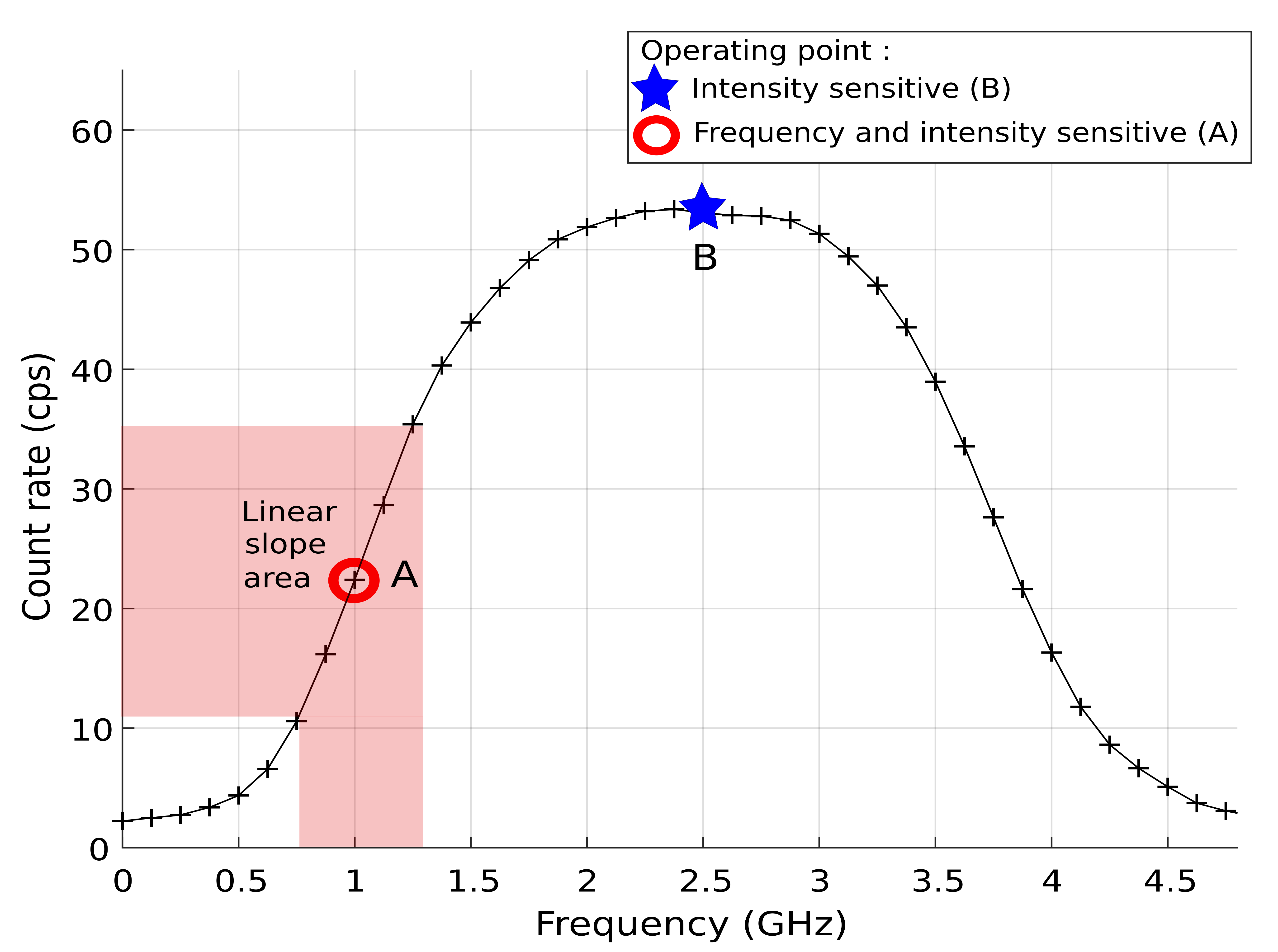}
\caption{Frequency response of our narrow FBG, scanned with SPAD, with an acquisition times equal to  60~s. Red circle: operating point A for frequency measurement. Blue star: operating point B, for intensity measurement only.}
\label{fig:FBG}
\end{figure}

In this article, we propose to use a frequency discriminator. This allows us to transform a frequency variation into an intensity variation. Thanks to this method, we do not need to make a frequency scan. By avoiding the frequency sweep, we also reduce the measurement time.  \\
Here, we propose to use the slope of a narrow FBG (red area on Fig.~\ref{fig:FBG}) to achieve this functionality and measure the Brillouin frequency. The FBG frequency response is shown in Fig.~\ref{fig:FBG}. From the frequency shift, we can deduce the temperature shift with the temperature sensitivity coefficient $C_{\nu_B T}$~=~1.07~MHz/K for a step index single-mode fiber at 1550~nm \cite{hartog_introduction_2017}.

It shows that, for an operating point halfway down the left-hand slope of the transmission band, we get 5\% transmission variation for 41~MHz frequency shift. This ratio $\eta_c$~=~8.4~MHz/\% enables to turn a frequency shift variation into a count rate variation on the SPAD. In order to perform distributed temperature measurement,  the propagation losses shown in Fig.~\ref{fig:Range} must be removed. Count rate datas are thus processed in order to compensate for the fiber losses, as a function of the probed distance, as depicted in Fig.~\ref{fig:Postprocessing_principle_bloc}. 

\begin{figure}[ht!]
\centering
\includegraphics[width=3.5 in]{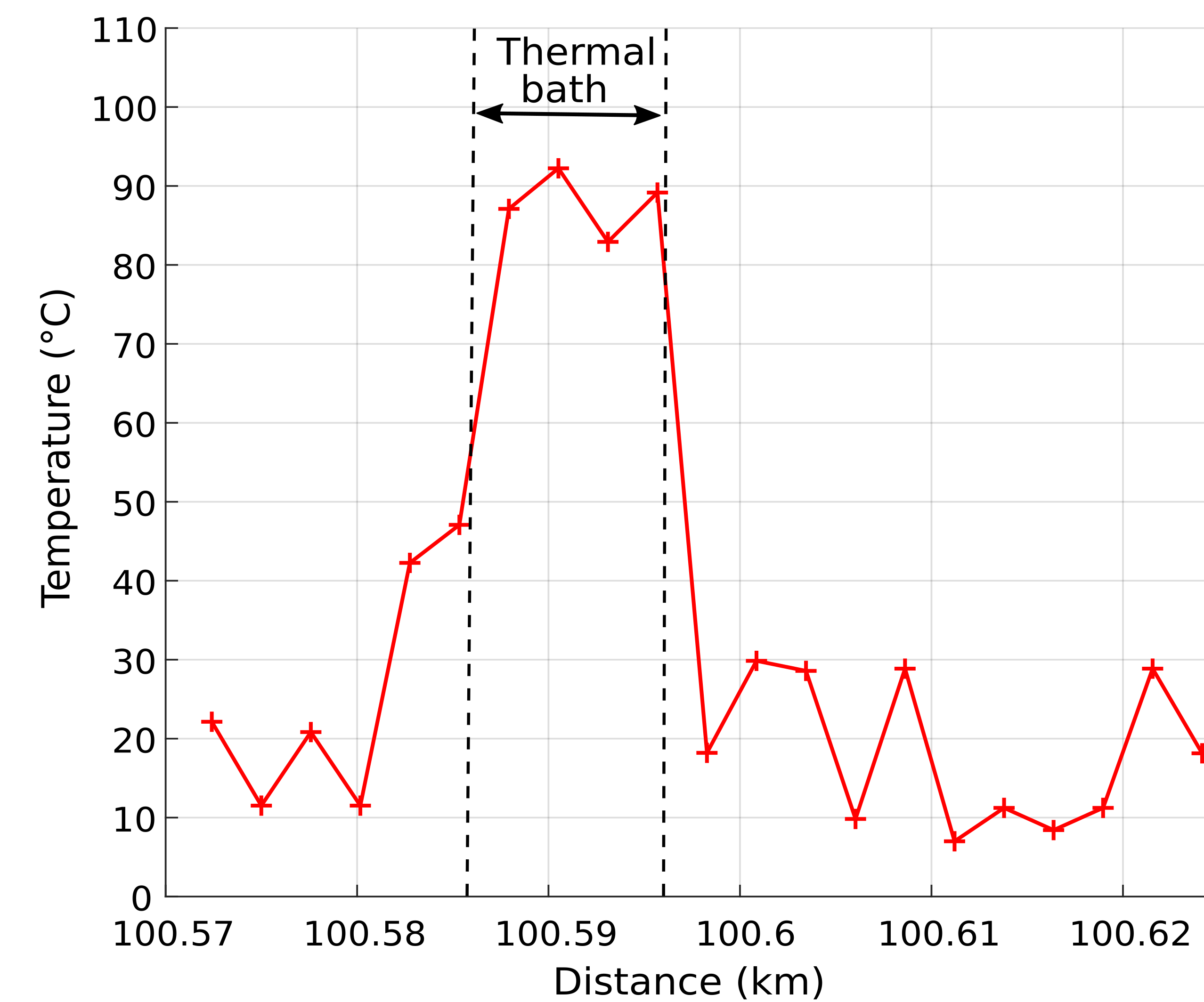}
\caption{Measured temperature profile between 100.57 km and 100.63 km, with 10~m fiber spool in bath at $86 \pm 1.5 ^\circ C$. Operating point A (see Fig.~\ref{fig:FBG}) was used with a 10~m spatial resolution and 10 minutes acquisition times.}
\label{fig:Temp}
\end{figure}

In Fig.~\ref{fig:Temp}, we report the temperature measurement obtained this way at operating point A (see Fig.~\ref{fig:FBG}) from a 10~m sample of a fiber immersed in $86 \pm 1.5~ ^\circ C$ water after 100~km. 

We demonstrate the ability to detect a hot point of 10~m, at 100~km, with 100~ns optical pulses. The profile is realized with a sampling interval of 2.5~m. In gated mode, the sampling interval corresponds to the time delay between two active gates on SPAD. At this distance, we get a count rate of 0.55~count per second (cps), corresponding to a SNR of 10.4~dB. 
The points measured at $45$~°C outside the bath simply correspond to samples overlapping the thermal transition, linked to a spatial resolution of 10~m. Moreover, the jitter delay line of 1.5~ns adds uncertainty to the sampling. \\

Even if we use the FBG slope to avoid a frequency scan, the acquisition time is quite long compared to conventional BOTDR. This is due to an acquisition point per point along the optical fiber which is intrinsic to the gated mode of the SPAD. A more detailed acquisition time measurement of our $\nu -$BOTDRr is a subject for future work. This measurement confirms the narrow FBG as a valid frequency discriminator, to convert count rate variation into temperature measurements. \\

\section{Real-time monitoring and data post processing}
\begin{figure}[ht!]
\centering
\includegraphics[width=3 in]{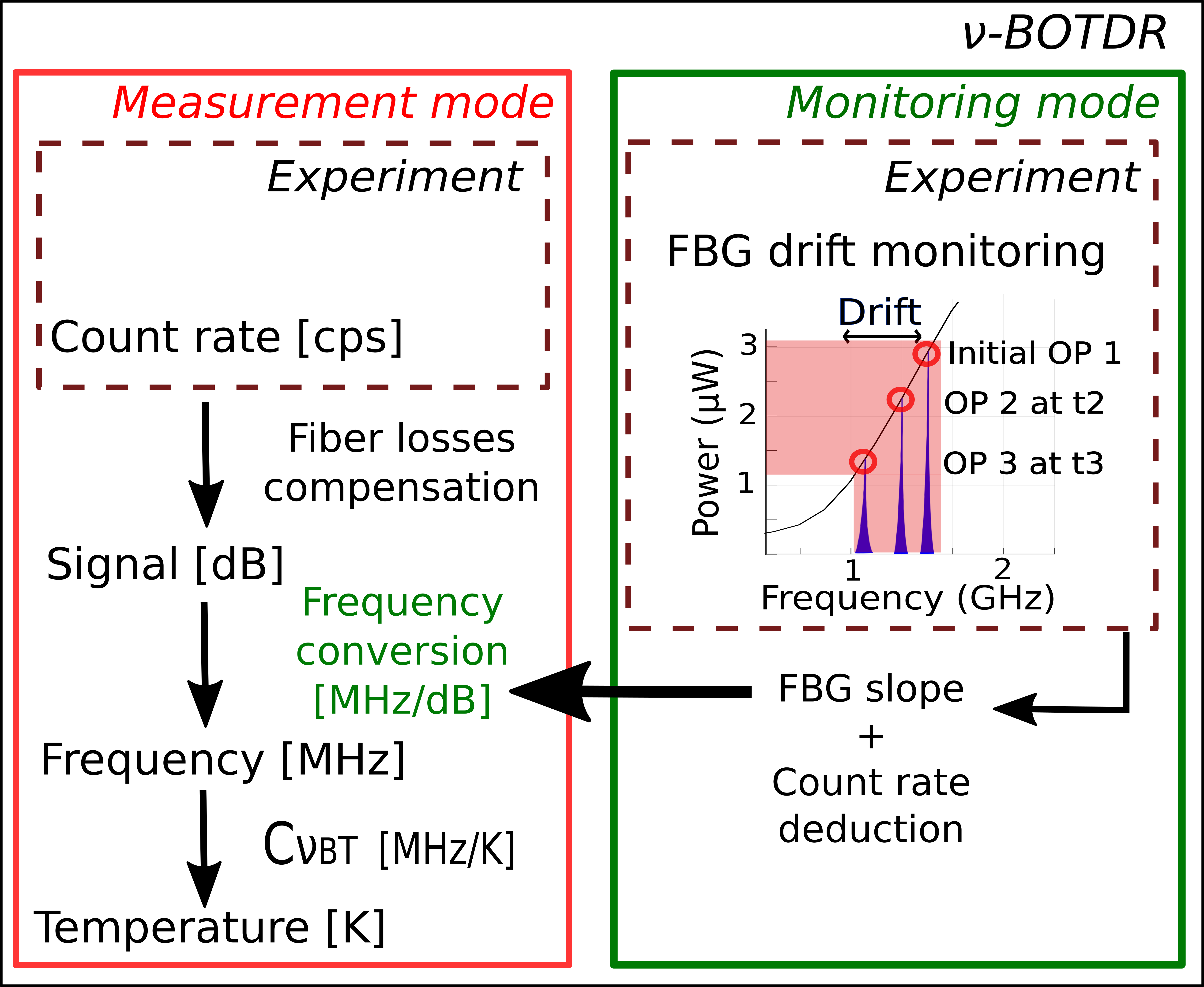}
\caption{Bloc diagram explaining the real-time monitoring of the $\nu-$BOTDR. In red, the data processing of the \textit{measurement mode} is explained. In green, the \textit{monitoring mode} is shown. It allows scanning the FBG shift every 10 minutes, in less than 1 minute, in order to compute the new operating point (OP) and the FBG slope. The Brillouin anti-Stokes resonance is represented in blue on the diagram. The contribution of the new operating point to the statistical measurement of the Brillouin frequency shift is corrected accordingly.}
\label{fig:Postprocessing_principle_bloc}
\end{figure}

The use of a frequency discriminator such as a narrow FBG has an obvious disadvantage. The relative frequency deviation between the laser and the FBG frequency is expected to be perfectly stable. Such setup will be sensitive to any form of instability (laser drift, high temperature variation). It thus usually requires a feedback loop to be reliable at the cost of increased complexity. However, this is not necessary in our case if we take advantage of the statistical nature of single-photon measurements. Indeed, the conversion of the count rate into Brillouin frequency shift is done by knowing the slope of the FBG at the given operating point. This operating point varies with the relative frequency difference between the laser and the FBG frequency. By measuring this difference, a new operating point can be deduced and the Brillouin frequency shift calculated with an updated conversion parameter $\eta_c$. Since single-photon detectors record events, it is only necessary to apply the measured relative frequency position of the FBG on experimental data in order to correct the drift as shown in Fig.~\ref{fig:Postprocessing_principle_bloc}. 

To achieve this, first we use a FBG with an athermal package in order to prevent thermal shocks and rapid temperature variations, leaving only slow temperature drifts ($>$10 min). 
Every 10 minutes, the setup switches to \textit{monitoring mode} and the relative deviation between the laser and the FBG is monitored  using the part of the setup highlighted in green in Fig.~\ref{fig:setup}. Switches SW1 and SW2 are then used to take some of the light from the laser and pass it through an EOM modulator. The modulator produces a frequency-adjustable harmonic that probes the FBG filter at 11~GHz. A frequency sweeping of the laser allows scanning the FBG, and the transmission is measured by the photodiode PD. Thus, FBG frequency shift with respect to the laser frequency is measured. The position of the operating point at 11 GHz of the laser frequency for a temperature of 300~K can be computed (red area on Fig.~\ref{fig:FBG}). If the FBG has drifted with respect to the laser, the slope at the operating point is measured and the following registered single-photon detections will be post processed accordingly as shown in green in Fig.~\ref{fig:Postprocessing_principle_bloc}. \\
Thanks to the FBG scanning, it is possible to compensate for filter drift, and thus to correct frequency measurements thanks to a dedicated post-processing. 

\section{Towards cross-sensitivity measurements}
\begin{figure}[ht!]
\centering
\includegraphics[width=3.5 in]{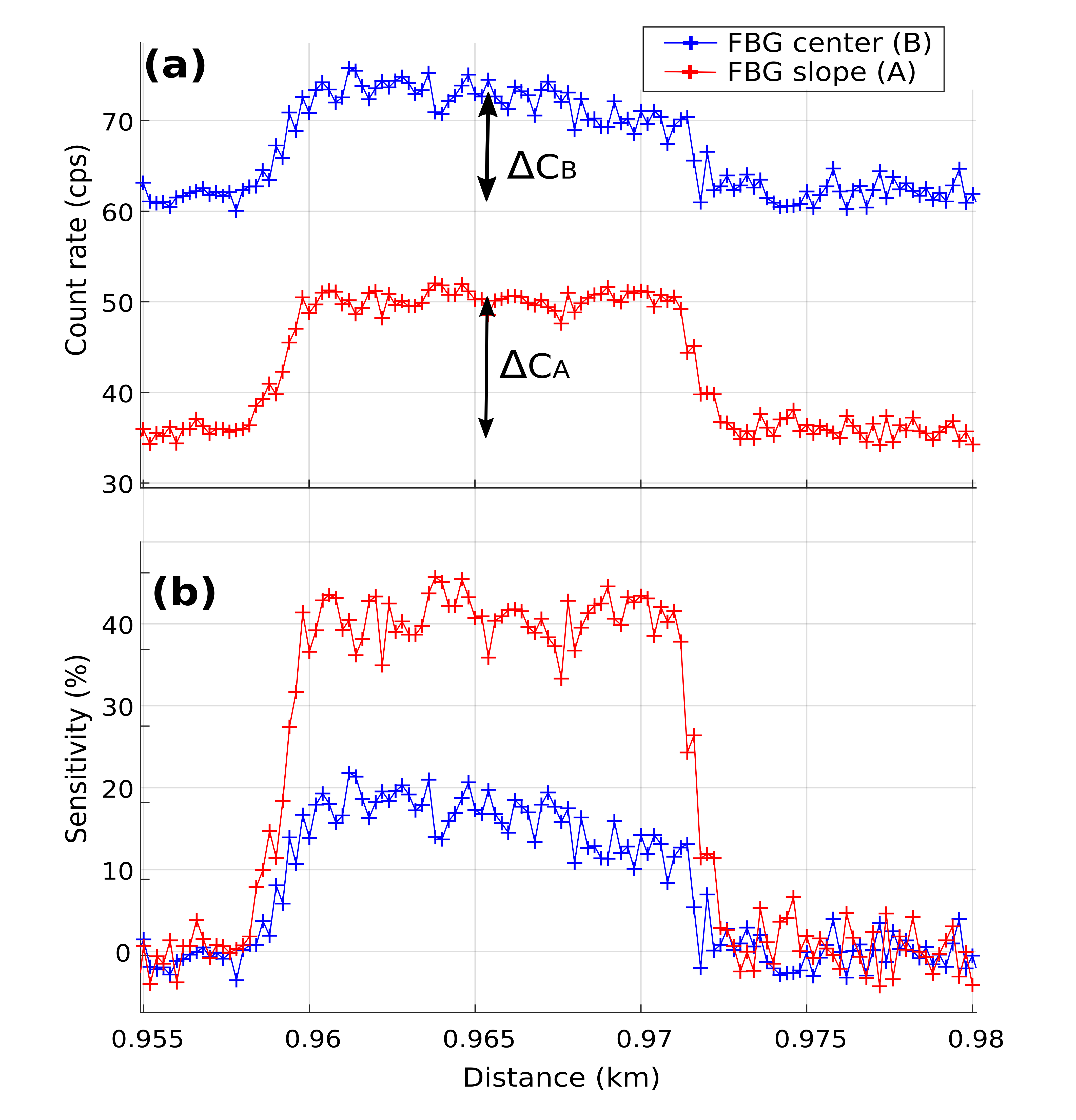}
\caption{Comparison of 2 working points A and B (see Fig.~\ref{fig:FBG}), with a 10~m fiber spool in bath at $84 \pm 1.5 ^\circ C$, at 1~km, with 15~ns optical pulses. Blue: Measurements realized at operating point B. Red: Measurements realized at operating point A. (a) Original data measurement. (b) Sensitivity comparison of both methods, with a reference at ambient temperature between 0.975~m and 0.980~m by averaging counts on this area.}
\label{fig: comparison_working_point}
\end{figure}

One of the caveats of our setup is that the count rate we measure on working point A (see Fig.~\ref{fig:FBG}) depends on both the Brillouin frequency shift and the efficiency of the Brillouin scattering. Both varies with temperature and strain \cite{hartog_introduction_2017}. One way to overcome this problem is to perform an additional control measurement when a temperature variation is recorded. By setting the laser frequency to change the operating point to $B$ in the middle of the FBG band pass filter as shown in Fig.~\ref{fig:FBG}, we only measure the efficiency of the Brillouin scattering. The superposition of both measurement are reported in Fig.~\ref{fig: comparison_working_point}, where the sensitivity is computed for the two methods. 

One can see in red the measurement with operating point A that includes both frequency shift and intensity contribution, while the measurement in blue only gives the intensity contribution. 
On Fig.~\ref{fig: comparison_working_point} we show the sensitivity of both methods, defined as: 
\begin{equation}
\centering 
Sensitivity = \frac{Count \; rate}{{\mathrm{avg}(Count \; rate_{Ambiant})}} \times 100 \quad [\%],  
\end{equation}
where $Count \; rate_{Ambiant}$ corresponds to the count rate averaging between 975 and 980~m, which is the reference level signal at room temperature. We clearly see that sensitivity at working point A, by using the slope of the FBG is higher ($41\%$) than the B working point ($16\%$) even if the count rate is lower for operational point A due to FBG losses. The highest sensitivity is obtained thanks to the frequency sensitivity ratio $\eta_c$ linked to the slope. Thereby, our method demonstrates a better sensitivity compared to LPR method based on Brillouin and Rayleigh intensity measurements \cite{xia_distributed_2016}. 
\\
Similarly to \cite{hartog_introduction_2017}, if we measure the count rate variation for a given temperature and strain variation for both A and B operating points, we can obtain the coefficients of the matrix M as seen on the right-hand side of
\begin{eqnarray}
  \label{strain_temperature_matrix}
		\left(
	\begin{array}{c}
	\Delta C_A\\
	\Delta C_B
	\end{array}
	\right)
	&=&
		\left(
	\begin{array}{cc}
	a_{\Delta T,A} & a_{\Delta T,B}\\
	a_{\Delta \epsilon,A} & a_{\Delta \epsilon,B}\\
	\end{array}
	\right)
	\left(
	\begin{array}{c}
	\Delta T\\
	\Delta\epsilon
	\end{array}
	\right),
	\label{eq_1}\\
	  &=&M	\left(
	\begin{array}{c}
	\Delta T\\
	\Delta\epsilon
	\end{array}
	\right). \nonumber
\end{eqnarray}

By inverting this matrix, one can retrieve the temperature and strain from both count rates variation.
\begin{eqnarray}
	\left(
	\begin{array}{c}
	\Delta T\\
	\Delta\epsilon
	\end{array}
	\right)
	&=&
	M^{-1}
		\left(
	\begin{array}{c}
	\Delta C_A\\
	\Delta C_B
	\end{array}
	\right)
	\label{eq_2}
 \label{strain_temperature_matrix_simplified}
\end{eqnarray}

 Where $\Delta C_A$ and $\Delta C_B$ are respectively the count rate variation measured for working point A and B. The temperature shift coefficients $a_{\Delta T,A}$, $a_{\Delta T,B}$ and strain shift variation $a_{\Delta \epsilon,A}$, $ a_{\Delta \epsilon,B}$ are related to temperature variation $\Delta T$ and strain variation $\Delta\epsilon$ with (\ref{strain_temperature_matrix}). 
 \\
 A measurement at the working point A and B would allow performing temperature and strain measurements separately.

\pagebreak

\section{Conclusion}
We have developed a $\nu-$BOTDR based on photon counting technology. The advantage of using SPADs for distributed sensor is the ability to measure small signal down to $-~120$~dBm, without the need for a cryostat, unlike SNSPDs. 
The high signal-to-noise ratio (SNR) allows measuring the distributed Brillouin response up to $120$~km with a spatial resolution of $10$~m in standard single mode fiber. We demonstrate the ability to detect a hot spot at $100$~km, and obtain a higher sensibility using a FBG compared to the LPR method. 
Thanks to the monitoring mode, we are able to correct environmental instabilities and calibrate our sensor. By using different operating points on the FBG we can optimize the measurement, and we show the possibility to discriminate strain and temperature effect without measurement of Rayleigh intensity.

\section*{Funding}
The authors would like to acknowledge the financial support of EIPHI Graduate School (contract ANR-17-EURE-0002), European ACTPHAST 4.0 project and French research ministry.

\section*{Acknowledgment}
We thank Marc Niklès and Félix Bussières for fruitful discussions, Jean-Marc Merolla for helpful discussion and AUREA technology for support in SPAD.
\ifCLASSOPTIONcaptionsoff
  \newpage
\fi

\bibliographystyle{IEEEtran}

\bibliography{Romanet1}

\end{document}